\documentclass[5p]{elsarticle}
\usepackage{amsmath}
\usepackage{epstopdf}

\usepackage{color}
\newcommand{\tr}{\mbox{ Tr}}

\newcommand{\Poly}{\mbox{Poly}}

\begin{document}

\begin{frontmatter}
\title{The transition to a layered phase in the anisotropic five-dimensional $SU(2)$ Yang-Mills theory}

\author{Luigi Del Debbio}
\ead{luigi.del.debbio@ed.ac.uk}

\author{Richard D. Kenway}
\ead{r.d.kenway@ed.ac.uk}

\author{Eliana Lambrou}
\ead{e.lambrou@ed.ac.uk}

\author{Enrico Rinaldi}
\ead{e.rinaldi@ed.ac.uk}

\address{SUPA, School of Physics and Astronomy, University of Edinburgh, Edinburgh EH9 3JZ, UK}

\begin{abstract}
We extend to large lattices the work of a previous investigation of the phase diagram of the anisotropic five-dimensional SU(2) Yang-Mills model using Monte Carlo simulations in the regime where the lattice spacing in the fifth dimension is larger than in the other four dimensions. We find a first order phase transition between the confining and deconfining phase at the anisotropic parameter point $\beta_4=2.60$ which was previously claimed to be the critical point at which the order of the transition changes from first to second. We conclude that large lattices are required to establish the first order nature of this line of transitions and consequently that the scenario of dimensional reduction of the five-dimensional theory to a continuum four-dimensional theory via the existence of the so-called ``layer phase" is unpromising. 
\end{abstract}

\begin{keyword}
Lattice Gauge Theories, Extra Dimensions, $\mbox{SU}(2)$ Yang-Mills Theory, Layer Phase
\end{keyword}

\end{frontmatter}

\section{Introduction}
Since the idea of introducing extra dimensions to give a better understanding of the hierarchy between weak and Planck scales was introduced more than a decade ago in~\cite{ArkaniHamed:1998nn}, many models have been developed to show how extra dimensional theories might solve these problems and yet exhibit a connection to the four-dimensional world that we observe. Well-known models that lead to dimensional reduction are those of Randall-Sundrum (RS)~\cite{RS, RS2}, ADD~\cite{ArkaniHamed:1998nn, ArkaniHamed:1998rs, Dienes}, Dvali-Shifman mechanism (DS)~\cite{DS} and compactification of the extra dimensions~\cite{Ejiri,Forcrand,Francesco,ER} and since the work of~\cite{Antoniadis:1990ew} they have gained phenomenological interest.  

The first three models envisage the four-dimensional world as layers that exist in the extra dimensions and are decoupled from each other. All the particles that we currently observe can only propagate in one layer, with the exception of the graviton that can interact between the layers. The RS model deals with warped space-time, while the ADD model (D-brane models) and DS mechanism consider the extra dimensions as flat. The compactification of the extra dimension is based on Kaluza and Klein's attempt to unify fundamental forces in the early 20th century~\cite{Kaluza, Klein}.  

Since higher-dimensional non-abelian models are non-renormalizable, it is necessary to define an effective theory by defining parameter values at which there is a separation of scales between the cutoff $\Lambda $ and low energy physics. The first study of the localization of fields on branes using non-perturbative methods came from Fu and Nielsen~\cite{Fu}. They propose that,  when there is a $D$-dimensional lattice, a $d$-dimensional layer phase  ($D = n+d$)  can be formed if the nearest-neighbour gauge couplings of the $d$-dimensional sublattice are different from the other $n$ dimensions. Then, particles and gauge fields can travel within the $d$-dimensional layer phase, but they exhibit a kind of confinement when they try to propagate in any of the $n$ extra dimensions.  The existence of the layer phase was shown in the five-dimensional U(1) gauge theory using Monte Carlo simulations in~\cite{Dimopoulos2, DimopoulosU1, U1_Coulomb}. 

When the lattice SU(2) Yang-Mills model is considered in five dimensions, the so called ``layer'' phase is believed to exist when the anisotropy in the lattice couplings is such that the lattice spacing in the extra dimension is larger than the one in the usual four dimensions, i.e. $a_5 > a_4$. In this layer phase one expects a zero string tension in the usual four-dimensional directions, but a non-zero string tension along the extra dimension.

The phase diagram of the five-dimensional non-abelian gauge theory was first studied in 1979 by Creutz~\cite{Creutz}, who showed a phase transition between the confined and the deconfined phase of the model with only isotropic couplings. During the last decade, the dimensional reduction from the five-dimensional theory to four dimensions, by implementing the model on a torus, was investigated on an anisotropic lattice using Monte Carlo techniques~\cite{Ejiri, Forcrand, Francesco, ER}. These papers show that, when the model is dimensionally reduced via compactification, the phase transition changes its nature to second order and belongs to the same universality class as the four-dimensional Ising Model. In~\cite{Ejiri, Forcrand, ER} simulations were done in the region where $a_4>a_5$, whereas in~\cite{Francesco} the simulations were carried out in the anisotropy region which is of interest in the case of the layer phase, i.e.  $a_4<a_5$. 

The existence of the layer phase was investigated by mean-field approximation on an anisotropic lattice with periodic boundary conditions~\cite{MeanField1, MeanField2} and it was shown that the planes transverse to the extra dimension were decoupled from each other. An investigation of the existence of this layer phase using Monte Carlo techniques was attempted by Farakos et al.~\cite{Farakos}. They claim that the transition between the five-dimensional Coulombic (deconfined) phase and the strong-coupling (confined) phase changes its order from first to second, implying that in the fifth dimension the layer phase exists. This opens a possibility of defining a continuum four-dimensional field theory.

The motivation for studying this model further came mostly from results in~\cite{Francesco}, which indicate that the lattice volumes used in the previous Monte Carlo simulations~\cite{Farakos}, were too small to show the correct order of the phase transition. In this letter we extend this investigation to larger volumes.   

This letter is structured as following: In Section 2 we set up the lattice model and the observables that were measured. In Section 3 we give details for our simulations and we present our results, and in Section 4 we give a brief conclusion.

\section{The Model}
\subsection{Anisotropic Action}
We investigate the anisotropic SU(2) Yang-Mills gauge theory in five dimensions, whose action in the continuum is given by
\begin{equation}\label{ActionCont}
S_E = \int d^4 x \int dx_5 \frac{1}{2g_5^2} \tr F^2_{MN}
\end{equation}
where $M,N = 1 \dots 5$ and $F_{MN} = \partial_MA_N - \partial_N A_M + i[A_M, A_N]$ with $A_M = g_5A_M^a T^a$.

On the lattice the action of the model becomes
\begin{equation}\label{WilsonAction}
\begin{split}
S = \beta_4 & \sum_x \sum_{1 \le\mu<\nu\le 4}\Big( 1 - \frac{1}{2} \tr\: U_{\mu\nu}(x) \Big)  \\
+ &\beta_5 \sum_x \sum_{1 \le\mu\le 4}\Big( 1 - \frac{1}{2} \tr\: U_{\mu 5}(x) \Big) \;\; _{\mu,\nu = 1 \ldots 4}
\end{split}
\end{equation}
where $U_{\mu\nu}(x)$ represents the oriented plaquette along spacetime directions given by
\begin{equation}
U_{\mu\nu}(x)= U_\mu(x)U_\nu(x+\hat\mu a_4)U^\dagger_\mu(x+\hat\nu a_4)U^\dagger_\nu(x)
\end{equation}
and $U_{\mu 5}(x)$ represents the plaquette formed when one of the directions is the extra-dimensional one, given by
\begin{equation}
U_{\mu 5}(x)= U_\mu(x)U_5(x+\hat\mu a_4)U^\dagger_\mu(x+\hat 5 a_5)U^\dagger_5(x).
\end{equation}
where $U_\mu = \exp(ig_5a_4 A_\mu)$ and $U_5=\exp(ig_5a_5A_5)$ are the gauge links and $a_4$ is the lattice spacing in the temporal and three spatial directions and $a_5$ is the lattice spacing in the extra direction.

The anisotropy parameter on the lattice is characterized by $\gamma$ which is given by
\begin{equation}\label{anisotropy}
\gamma = \sqrt{\frac{\beta_5}{\beta_4}}
\end{equation}
and at classical level this is given by 
\begin{equation} 
\gamma = \frac{a_4}{a_5}
\end{equation}

\subsection{Observables}
In order to investigate the phase diagram of the model we use the following observables:
\begin{itemize}
\item{Average Plaquette in the extra dimension $x_5$, $\hat P_5$ \\
\begin{equation}\label{Plaq5}
\langle \hat P_5 \rangle = \Big \langle \frac{1}{4VN_c} \sum_x \sum_{\mu} \mbox{Tr}(U_{\mu 5}(x)) \Big \rangle
\end{equation} \\
and its susceptibility
\begin{equation}
\chi_{\hat P_5}= V \Big ( \langle \hat P_5^2 \rangle - \langle \hat P_5 \rangle^2  \Big)
\end{equation}
where $V$ is lattice volume given by $V=L_T\times L_S^3 \times L_5$, with $L_T$, $L_S$ and $L_5$ the size of the temporal, spatial and extra dimension respectively.}
\item{Temporal Polyakov Loop. This can be measured on the whole lattice given by
\begin{equation} \label{Poly}
\Poly_T = \frac{L_T}{N_cV} \bigg |\sum_{\vec{x},x_5}\tr \prod ^{(L_T-1)a_4}_{x_1=0} U_1(x) \bigg |
\end{equation}
Since, in the layer phase, each layer is uncorrelated, the Polyakov loop may be measured in one layer and so we also compute   
\begin{equation} \label{PolySlice}
\Poly_T(x_5) = \frac{1}{N_cL_S^3} \bigg |\sum_{\vec{x}}\tr \prod ^{(L_T-1)a_4}_{x_1=0} U_1(x)|_{x_5} \bigg |
\end{equation}
We define the Polyakov loop susceptibilities as
\begin{equation}
\chi_{\Poly_T} = \frac{V}{L_T} \Big \langle \big( \Poly_T^2 - \langle\Poly_T\rangle^2 \big ) \Big \rangle
\end{equation}
\begin{equation}\label{eq:poly_x5slice}
\chi_{\Poly_T}(x_5) = L_S^3 \Big \langle \big( \Poly_T(x_5)^2 - \langle\Poly_T(x_5)\rangle^2 \big ) \Big \rangle.
\end{equation}
}
\end{itemize}
The expected behaviour of the plaquette for a first order phase transition is to show hysteresis in the expectation value and a divergence in the susceptibility at the critical point. The temporal Polyakov loop is expected to have a zero expectation value in the strong phase, i.e. to fluctuate around zero and a non-zero expectation value in the deconfining phase, i.e. to show a two-peak structure.    

\section{Results from Lattice Simulations}
Our model was implemented on the lattice, using the Kennedy-Pendleton Heat-Bath algorithm~\cite{KPHeatBath} combined with overrelaxation updates~\cite{Overrelaxation}. Specifically, we took one heat-bath measurement every $L_S/2$ overrelaxation steps. The autocorrelation that arises was taken into account in our analysis. The number of measurements varied between 100,000 and 200,000 at each set of points in our parameter space $(\beta_4, \beta_5)$ that were investigated. Measurements were taken starting from either random SU(2) matrices (hot configurations) or by setting all the SU(2) matrices to the identity matrix (cold configuration). The lattice volumes were $16^5$, $20^4\times8$ and $24^4\times8$. The former was the largest investigated by Farakos and Vrentzos~\cite{Farakos} and was included here as a check with their results. The bigger volumes were used, since~\cite{Francesco} showed that there is a minimum size of the spatial/temporal and the extra direction in order to see a clear first order phase transition. Since, $16^5$ is below this minimum size, we simulated bigger volumes to investigate the order of the phase transition. We reduced the size of the extra dimension to $L_5=8$ to save compute time, since the lattice spacing in the extra dimension is much larger for high $\beta_4$ values than the lattice spacing in the other directions (as shown in~\cite{Francesco}), so the system remains five dimensional.

\begin{figure}[!ht]
\centering
\includegraphics[scale=0.35, angle = 270]{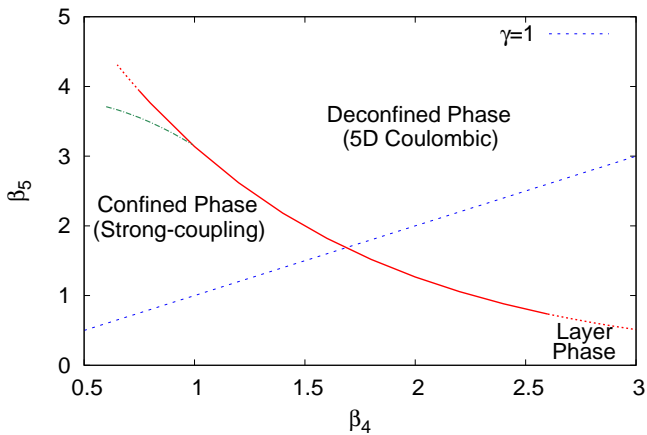}
\caption{A sketch of the phase diagram of the anisotropic SU(2) Yang-Mills model. The dashed blue line denotes the isotropic case $\gamma=1$. The region above this line was previously investigated in~\cite{Ejiri, Forcrand, ER} and the region below in~\cite{Francesco, Farakos}. The dashed-dotted green line appears when the extra dimension is compactified~\cite{Ejiri, Francesco, ER}. When no compactification is involved, there is a bulk phase transition which is shown in the figure as a red solid line. It was shown to exist up to $\beta_4=2.50$ in~\cite{Francesco}. In this work we extend the range of this line up to $\beta_4=2.60$ with no evidence that this line will not continue for larger values of $\beta_4$. For $\beta_4>2.60$ the idea of the existence of the layer phase arises.}
\label{fig:phase_diagram}
\end{figure}

\begin{figure}[!ht]
\centering
\includegraphics[angle=270, scale=0.35]{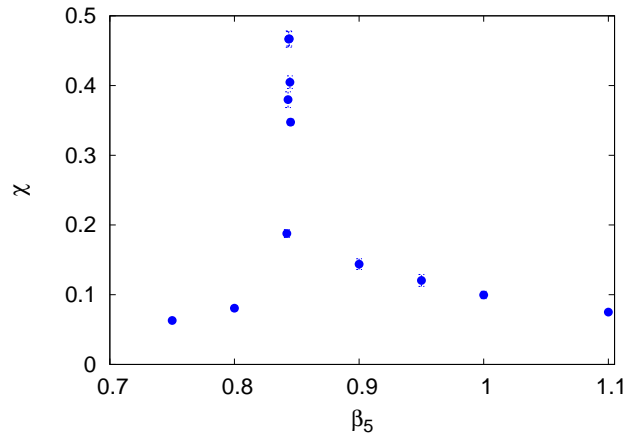}
\caption{The susceptibility of the plaquette in the extra dimension, $\hat P_5$ for $V=16^5$ keeping $\beta_4$ fixed at 2.60 and varying $\beta_5$. The critical point is the point at which the susceptibility gains its maximum value. }
\label{fig:Susc_V16_b4_260}
\end{figure}

First, we did a scan in the parameter space ($\beta_4,\beta_5$) using small lattices to identify the first order phase transition that was shown in previous work up to $\beta_5=2.50$. The phase diagram of the model is shown schematically in Fig. \ref{fig:phase_diagram}. The layer phase was previously claimed to exist at large $\beta_4$ and small $\beta_5$,  as shown. Our point of interest is $\beta_4=2.60$ on the line of transition, which was claimed to be the critical point at which the transition changes from first to second order in~\cite{Farakos}. The critical point in $\beta_5$ was found by implementing the model on a lattice of volume $16^5$. At this volume we were able to do  a wide scan by investigating a sufficient number of different $\beta_5$  to identify the critical point.  Even though it does not show any clear evidence of first order phase transition in terms of a two-state signal, by looking at the susceptibility it looks like it has a divergence at the critical point (Fig. \ref{fig:Susc_V16_b4_260}). The critical $\beta_5$ point was found to be $\beta_5=0.8437(5)$ which agrees within error with the value found in~\cite{Farakos}.

\begin{figure}[!ht] 
\centering
\includegraphics[angle=270, scale=0.17]{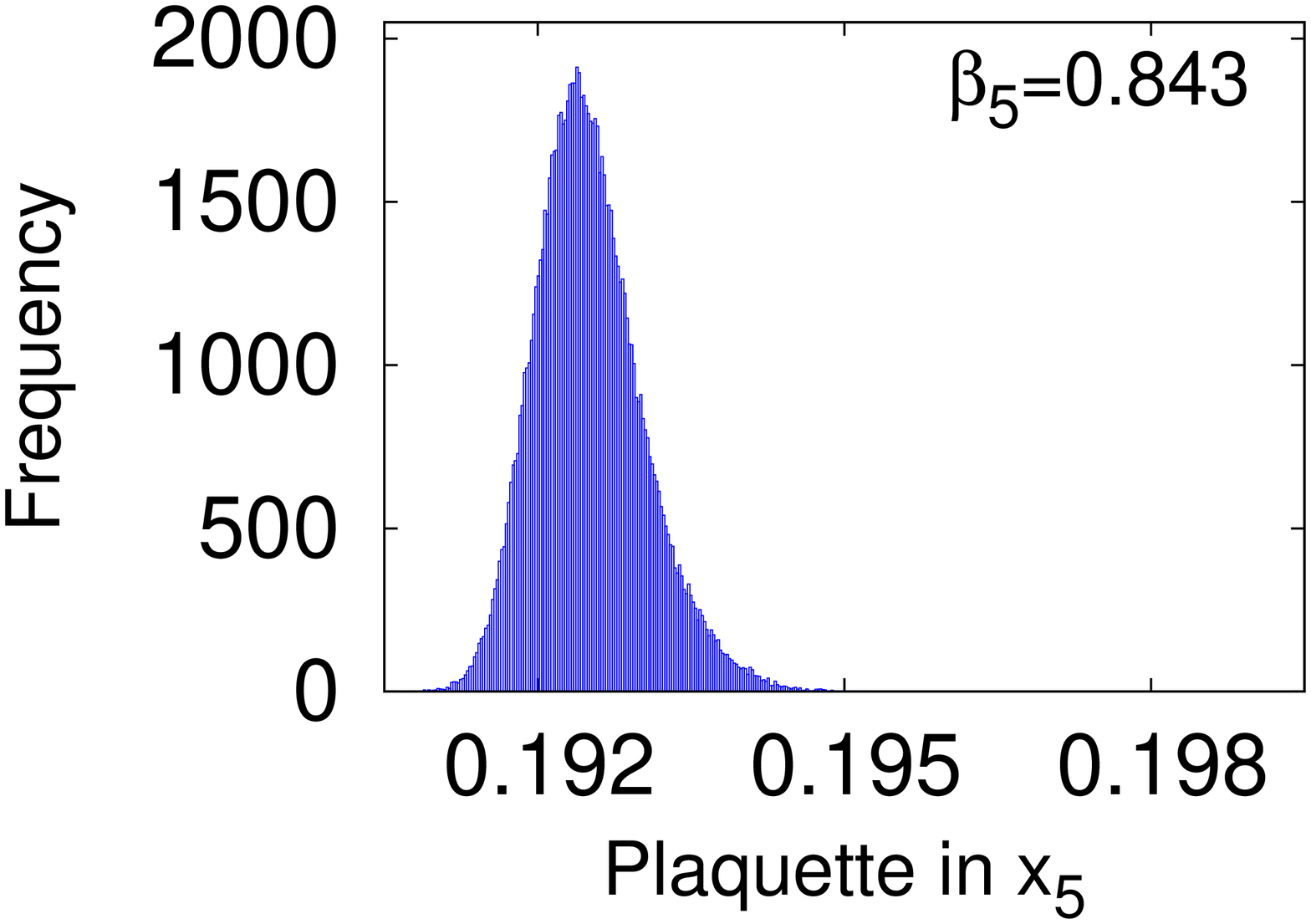}
\includegraphics[angle=270, scale=0.17]{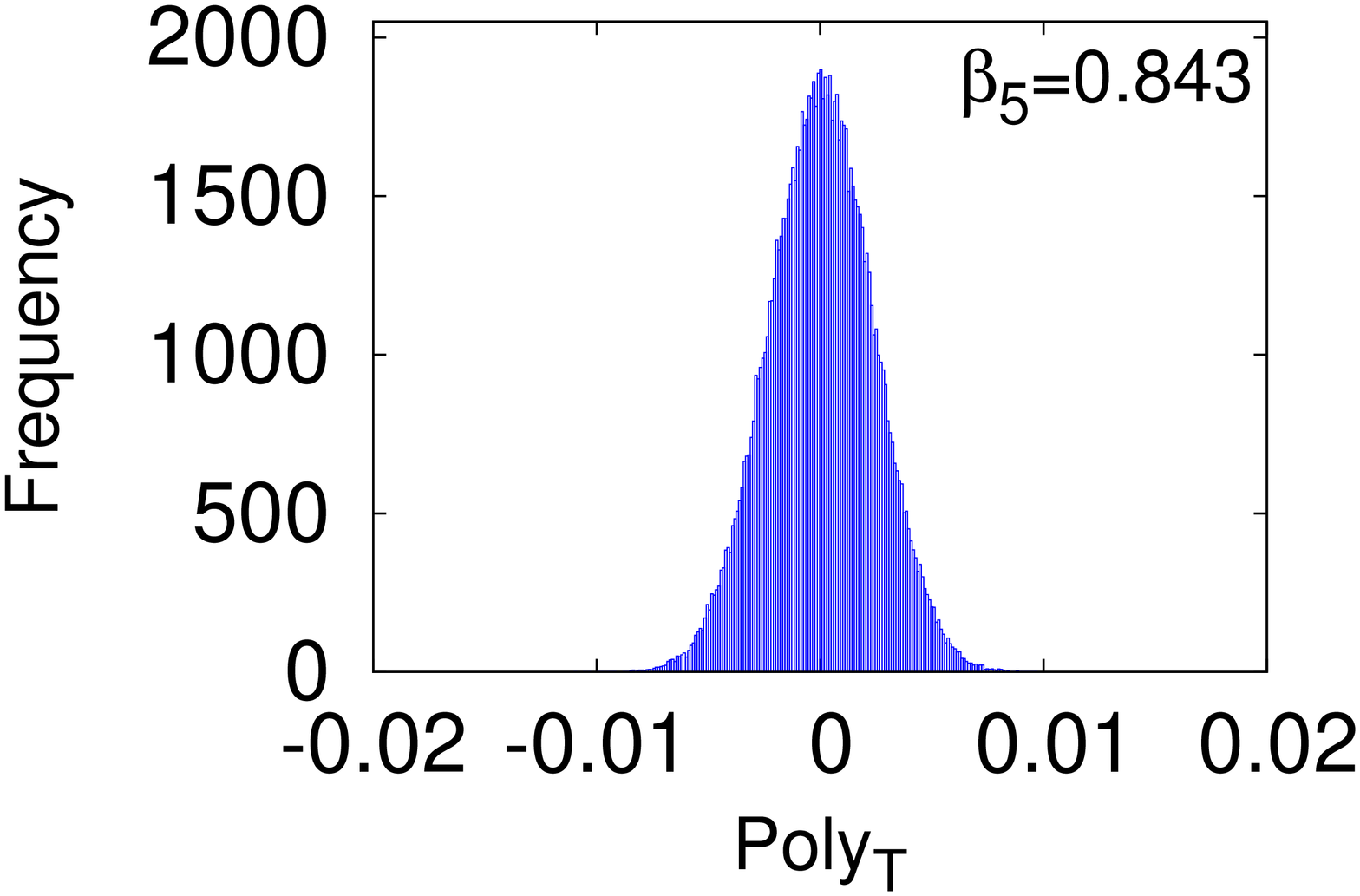}
\includegraphics[angle=270, scale=0.17]{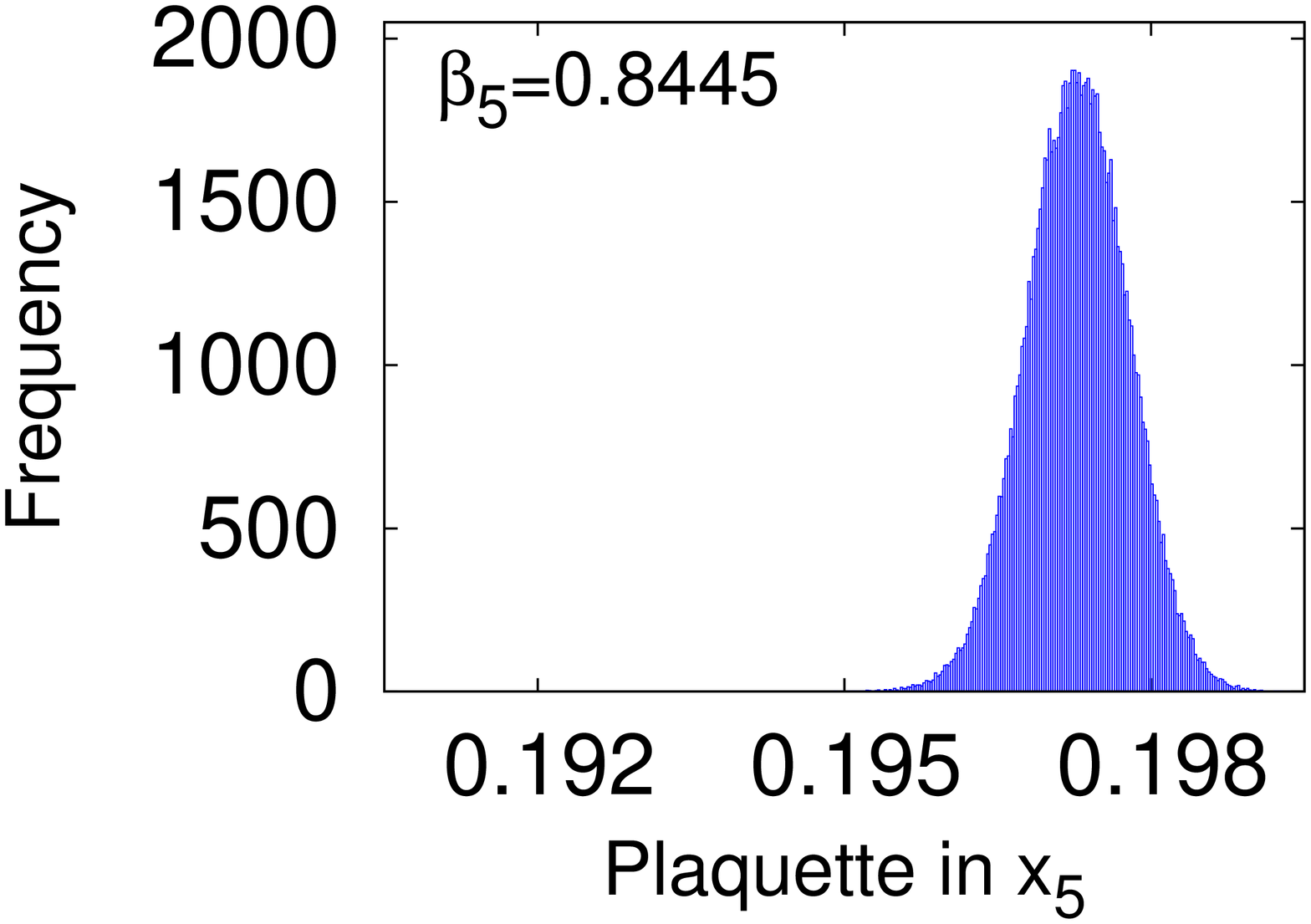}
\includegraphics[angle=270, scale=0.17]{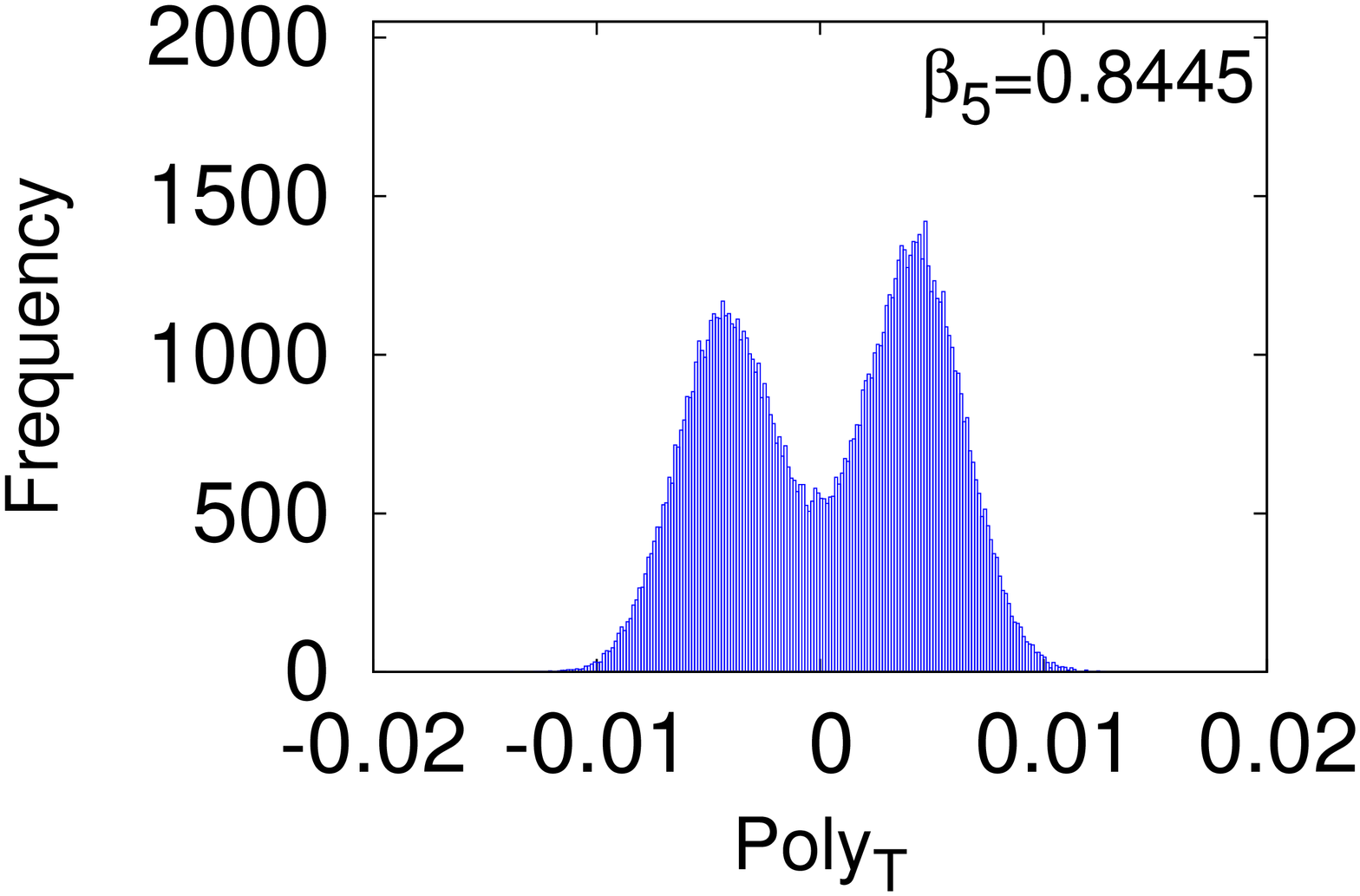}
\label{fig:plaq_and_poly0_V20_marginal}
\caption{The histograms for the plaquette in the extra dimension for $V=20^4\times8$, fixed $\beta_4=2.60$ and for two different values of $\beta_5$: $\beta_5=0.843$(top)  and $\beta_5=0.8445$(bottom) are shown on the left. We can see that the peak moves towards the right as we go to higher values of $\beta_5$. The corresponding histograms for the temporal Polaykov loop are shown on the right. We can see that for $\beta_5=0.843$ it has a zero expectation value whereas for $\beta_5=0.8445$ it shows a two-peak structure.}
\end{figure}

\begin{figure}[!ht]
\centering
\includegraphics[angle=270, scale=0.3]{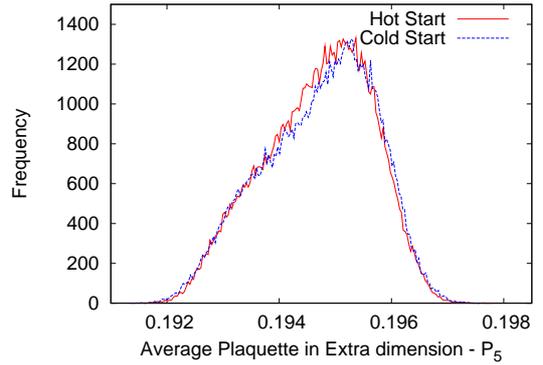}
\caption{Histograms of the average plaquette in the extra dimension, $\hat P_5$ starting from both cold and hot configurations for $V=20^4\times8$, $\beta_4=2.60$ and $\beta_5=0.8435$. We can see that since this point is very close to the critical one, the plaquette fluctuates between the two vacua and thus the distribution is not Gaussian anymore.}
 \label{fig:hist_V20_b4_260_b5_08435}
\end{figure}

\begin{figure}[!ht] 
\centering
\includegraphics[angle=270, scale=0.3]{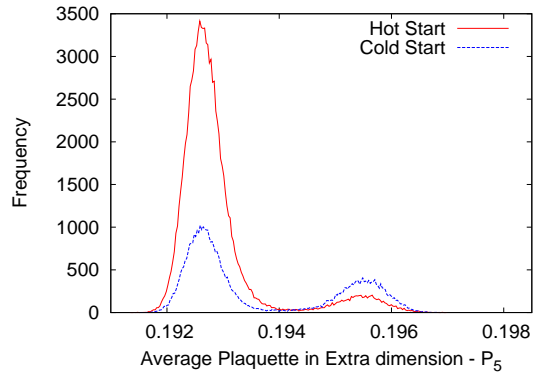}
\caption{Histograms of the average plaquette in the extra dimension, $\hat P_5$ starting from both cold and hot configurations for $V=24^4\times8$, $\beta_4=2.60$ and $\beta_5=0.8435$. Here, we can see that the distributions, starting from either cold or hot, build up as two Gaussian distributions, one for each vacuum that the system equilibrates to.}
\label{fig:hist_V24_b4_260_b5_08435}
\end{figure}

For the investigation of the phase transition on the larger lattices we focused on the critical region which was estimated to be between $\beta_5=0.843$ and $\beta_5=0.8445$, based on the critical value found for the $16^5$ lattice. As can be seen from Fig. \ref{fig:plaq_and_poly0_V20_marginal}, the plaquette moves to the right as we go to higher values of $\beta_5$ and the temporal Polyakov loop is zero for the point $\beta_5=0.843$, which is in the confining phase and has a two-peak structure for the point $\beta_5=0.8445$, which is in the deconfining phase, as expected. We also checked how the temporal Polyakov loop is distributed when all the $x_5$-slices are considered independently (Eq. \ref{PolySlice}) and we could see that each fluctuates around zero. Also, we confirmed that the critical point was included in this region by observing that, for one point that lies in between these values, either a clear two-state signal or large fluctuations between two values in the average value of the plaquette were present, as can be seen in Figures \ref{fig:hist_V20_b4_260_b5_08435} and \ref{fig:hist_V24_b4_260_b5_08435}. 

\begin{figure}[!ht]
\centering
\includegraphics[angle=270, scale=0.3]{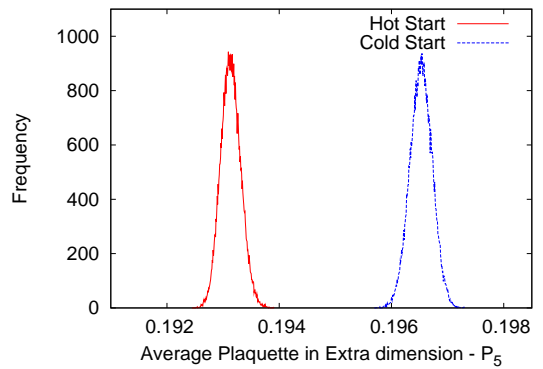}
\caption{Histograms of the average plaquette in the extra dimension, $\hat P_5$ starting from both cold and hot configurations for $V=32^4\times8$, $\beta_4=2.60$ and $\beta_5=0.844$. It is clear that a first order phase transition is present since starting from different configurations, the system equilibrates in different states with no tunnelling between them.}
\label{fig:hist_V32_b4_260_b5_0844}
\end{figure}

The points that were investigated are $\beta_5=0.843$, $0.8435$, $0.844$, $0.8445$. For the lattice volume of $20^4\times8$, we cannot distinguish the two states, since by starting either from hot or cold configurations, the expectation of the plaquette is the same. However, as shown in Fig.~\ref{fig:hist_V20_b4_260_b5_08435}, the distribution is not Gaussian. This is the first hint of the existence of a first order phase transition. The larger volume of $24^4\times8$ shows a clear two-state signal as can be seen from the histogram in Fig.~\ref{fig:hist_V24_b4_260_b5_08435}. The overlap that appears between the histograms starting from hot or cold configurations is due to the fact that the two vacua of the potential energy are not so deep and thus the system fluctuates between them. For a check, we implemented one single point ($\beta_5=0.844$) on a $32^4\times8$ lattice and we can see in Fig. \ref{fig:hist_V32_b4_260_b5_0844} that the two states are now separated by a wide gap, and there is no evidence of tunnelling from one to the other, i.e. it stays in the phase in which it first equilibrated,  depending on the initial configuration. This is also an indication that the extrapolation to the thermodynamic limit must be based on sufficiently large lattices.     

We note that the critical point was not estimated precisely, because reweighting techniques were not trustworthy for the large volumes and ensemble sizes that were used in this work due to limited statistics.

The code was written using QDP++~\cite{Chroma} and run on GPUs using QDP-JIT~\cite{Frank}. An estimate of the compute time required for the investigation of a single point on an NVIDIA Tesla C2070 Computing Processor (GPU) for the volumes used in this work for a set of 100,000 measurements  with $L_S/2$ overrelaxation steps and a heatbath update each time is shown in Table~\ref{table:GPU_compute_time}. The GPUs were provided by the Particle Physics Theory Group at the University of Edinburgh and the Edinburgh Compute and Data Facility. The single point of $V=32^4\times8$ would have taken two months on GPUs and so was simulated using STFCs DiRAC facilities in Edinburgh.  

\begin{table} [!ht]
\begin{center}
\begin{tabular} {|c|c|}
\hline
Lattice Volume & Compute time (hours) \\
\hline
$16\times16\times16\times16\times16$ & 190 \\
$20\times20\times20\times20\times8$ & 250 \\
$24\times24\times24\times24\times8$ & 620 \\
\hline
\end{tabular}
\caption{Estimated compute time required on an NVIDIA Tesla C2070 Computing Processor for 100,000 measurements for a single point in the parameter space $(\beta_4, \beta_5)$.}
\label{table:GPU_compute_time}
\end{center}
\end{table}
\section{Conclusions}
In this work, we extended the Monte Carlo investigation of the phase diagram of the anisotropic SU(2) Yang Mills model in five dimensions when the lattice spacing in the extra dimension is larger than that in the four other dimensions ($\gamma<1$). We showed that, up to $\beta_4=2.60$, there is no evidence of a second order phase transition, whereas a clear two-state signal in the average plaquette favours a first order phase transition. Based on this result, we can claim that the bulk first order phase transition between the confining and the deconfining phase continues at least up to $\beta_4=2.60$ with no trace of a layer phase. This implies that up to this point  the continuum limit cannot be taken and thus the possibility of a dimensionally reduced five-dimensional effective field theory remains open. Even though, based on the work of this letter, we cannot exclude a second order transition at higher $\beta_4$, nothing in our study suggests that continuing this investigation on even bigger lattices would be worthwhile. 
  
\section*{Acknowledgements}
\thispagestyle{empty}
\noindent L.D.D. and R.D.K. are supported by STFC Consolidated Grant ST/J000329/1 and by the EU under Grant Agreement PITN-GA-2009-238353 (ITN STRONGnet). E.L. is supported by an STFC studentship and E.R. is funded by a SUPA Prize Studentship. The DiRAC facilities used are supported by STFC grants ST/H008845/1, ST/K000411/1 and ST/K005804/1. We would like to thank Frank Winter for his support and guidelines on running the code on GPUs and Philippe De Forcrand, Antonio Rago, Francesco Knechtli and Nikos Irges for useful discussions during the implementation of this work.  

\bibliographystyle{h-elsevier3}
                 
\bibliography{references}    

\end{document}